# Polyrotaxane: New Generation of Sustainable, Ultra-flexible, Form-stable and Smart Phase Change Materials

Guang-Zhong Yin, Jose Hobson, Yanyan Duan, De-Yi Wang*
*IMDEA Materials Institute, C/Eric Kandel, 2, 28906 Getafe, Madrid, Spain*

**Abstract**
The development of thermal energy storage materials is the most attractive strategy to harvest the solar energy and increase the energy utilization efficiency. Phase change materials (PCMs) have received much attention in this research field for several decades. Herein, we reported a new kind of PCM micro topological structure, design direction, and the ultra-flexible, form-stable and smart PCMs, polyrotaxane. The structure of polyrotaxane was fully confirmed by $^1$H nuclear magnetic resonance, attenuated total reflection-fourier transform infrared and X-ray diffraction. Then the tensile properties, thermal stability in the air, phase change energy storage and shape memory properties of the films were systematically analyzed. The results showed that all the mechanical performance, thermal stability in air and shape memory properties of polyrotaxanes were enhanced significantly compared to those of polyethylene oxide (PEO). The form stability at temperatures above the melting point of PEO significantly increased with the α-CD addition. Further with the high phase transition enthalpy and excellent cycle performance, the polyrotaxane films are therefore promising sustainable and advanced form-stable phase change materials for thermal energy storage. Notably, its ultra-high flexibility, remolding ability and excellent shape memory properties provide a convenient way for the intelligent heat treatment packaging of complex and flexible electronic devices. In addition, this is a totally novel insight for polyrotaxane application and new design method for form-stable PCMs.

# 1. Introduction

The deterioration of fossil energy and the increase in environmental pollution have made the exploitation of clean, sustainable, and renewable energy resources increasingly desirable and challenging. [1] The development of thermal energy storage materials is the most attractive strategy to harvest the solar energy and increase the energy utilization efficiency. Phase change materials (PCMs) have received much attention in this research field because of their large thermal energy storage density, wide temperature working range, long-term stability, noncorrosive, and low toxicity properties.[2, 3] However, the commonly occurring issues of organic PCMs are their poor flexibility, complex manufacturing process, low thermal conductivity, and leakage during the solid-liquid phase-change process. [4, 5] As a credible alternative to solve this problem, some porous matrixes have usually been incorporated with organic PCMs to improve their thermal conductivity as well as form stability.

Polyethylene glycol (PEG)-based composite PCMs are widely employed as energy storage materials in the fields of solar energy utilization and waste heat recovery, due to their excellent thermal properties, large phase transformation enthalpy combined with suitable phase change temperature, non-toxicity, low cost, and biodegradability. [6] Moreover, the properties of PEG can be further manipulated via the adjustment of its molecular weight in order to optimize the performance. Typically for the leakage of PEG caused by the solid-liquid transformation is also the main hindrance for its practical application. Currently, there are three typical methods to generate form stable PEG composites:[7] (1) encapsulation of PEG in shell materials, [8] (2) preparation of polymer/PEG composites, [9] and (3) impregnation of PEG into inorganic materials with porous or layered structures, such as MXene, [10] graphene oxide, [11] graphene, [12] diatomite, [13] 3D ceramics or carbon network, [14, 15] and silica. [16] The first two methods have some disadvantages such as a low thermal conductivity, incongruent melting and freezing, latent heat capacities, [17] complex manufacturing processing, [18] and high cost accordingly. Furthermore, it is difficult to obtain flexible phase change energy storage materials because most of the prepared materials are rigid or powder shape, which require secondary processing for practical applications. Although some monolithic composite PCMs have been developed, their flexibility usually undergoes a remarkable reduction or even complete disappearance when supporting materials are infiltrated with PCMs.

More recently, PCMs have been widely concerned in the selection of heat dissipation materials for 5G electronic products and base stations due to their excellent heat absorption, heat storage capacity, no external driving force, and no noise. With the commercial application of 5G technology, the market of PCMs for thermal adjustment on the base station, mobile phone, flat panel computer and other mobile terminals is further opened. In addition, energy storage and electronic devices are developing towards flexible, lightweight, intelligent, and wearable, which requires high mechanical strength and flexibility of materials. Poly (ethylene oxide) (PEO) has also been widely used in chemically modified PEOs and polymer/PEO blends exhibit unique solid–solid phase transition behavior and are expected to be efficient flexible thermal energy storage materials.[19] However, the low Young's modulus and tensile strength of PEO films limit their applications. The utility of environment-friendly materials is currently considered as a necessity imposed by nature for securing a sustainable future against problems such as depletion of petrochemical resources and white pollution. [20] In this respect, developing green energy management systems is vital for green technological innovation and efficient energy utilization.[21]

To sum up, we hope to design and prepare new PCMs that (1) can be prepared in a facile and green pathway, and (2) have high flexibility, high strength, high form stability, and high phase transition enthalpy. Fortunately, polyrotaxanes (PLR) consisting of biomass cyclodextrin (CD) "wheels" and polymer "axles" is appealing mechanically supramolecular polymers. [22, 23] To solve the current problems of the conventional PCMs and fulfil the high-performance requirements for flexible devices, in this work, we will prepare high molecular PLRs with different α-CD contents by one facile method. The optical, mechanical properties (flexibility), thermal properties, form stability, and the heat response of the PLRs will be investigated as a function of the α-CD contents for application as high-performance form-stable PCMs.

## 2. Results and discussion
*2.1 Synthesis and structure characterization of PLR*

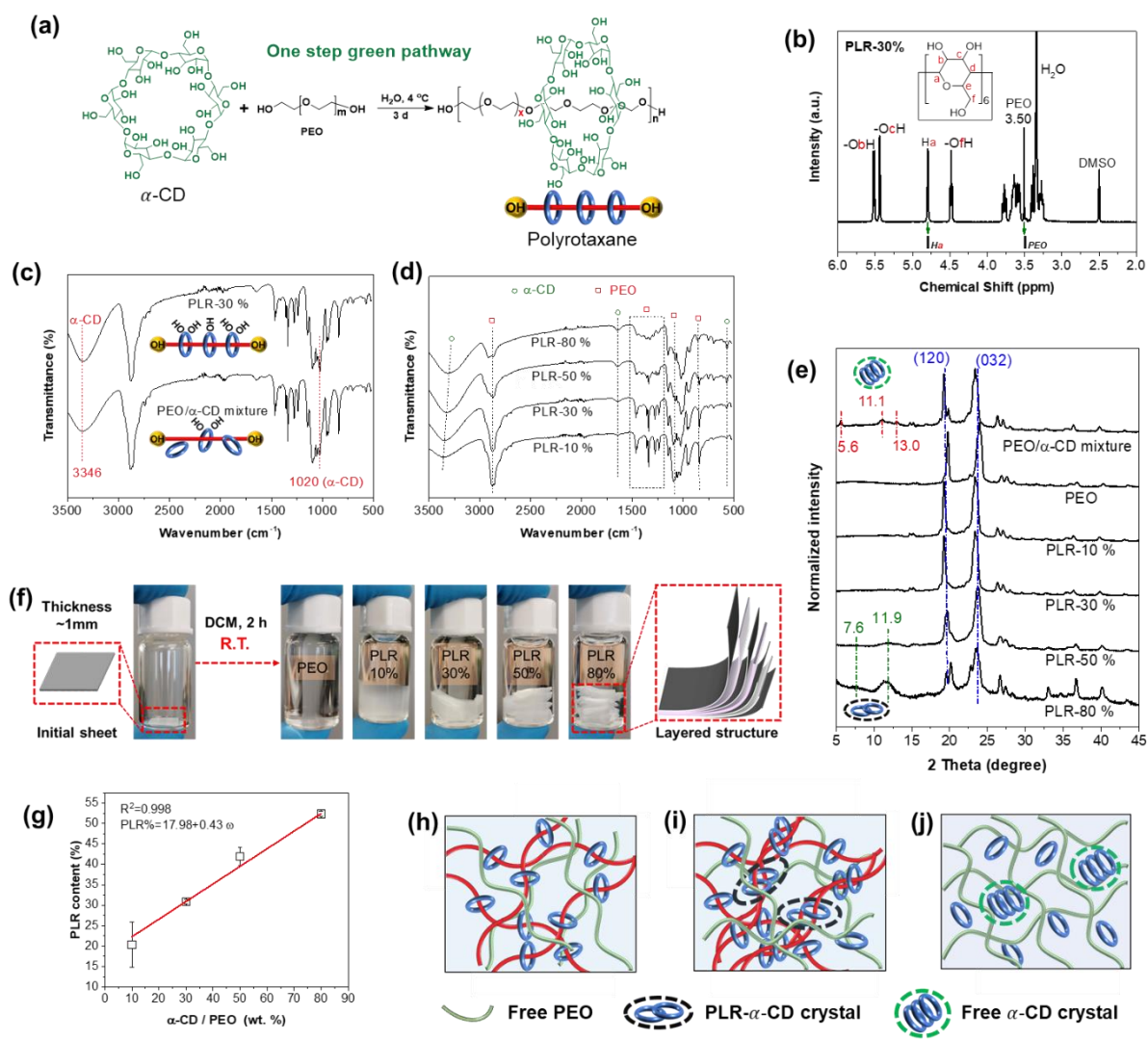

**Figure 1**. (a) PLR synthesis route, (b) $^1$H NMR curve of sample PLR-30 % as a typical example, (c) FTIR spectra of PLR-30 % and PEO/α-CD mixture (30 wt.%), (d) FTIR spectra of all the four PLRs, (e) XRD curves of the four PLR samples, PEO and PEO/α-CD mixture, (f) solvent resistance test: Comparison of sample morphology before and after solution treatment, and samples in DCM at room temperature (RT) for 2 h, (g) the measured contents of the solid that can't be removed by Dichloromethane treatments, and micro scale structure illustration of (h) PLRs with low α-CD content, (i) PLR with high α-CD content and (j) PEO/α-CD mixture.

The well reported PLRs (PEG/α-CD) were widely used in drug loading, hydrogel toughening, molecular machines and other intelligent or biomedical fields. [24] [22, 25, 26] Generally, PEG (or PEO) based PLRs are subject to end-capping modification. The purpose of end-capping is to prevent the separation between *α*-CD and PEO during subsequent process in the aqueous phase. In this work, we did not perform end-capping and chemical modification in order to avoid the two-step chemical reactions for the end capping, dialysis purification and the freeze-drying step. Because the synthesis method is simple, and only water is used as a solvent without high temperature treatment, we assigned the method here a totally green synthesis method. The experimental part is provided in Supporting Information. Moreover, the one-step high-efficiency synthesis will lay a solid foundation for the practical application of the PLRs.

**Figure 1a** shows the synthesis route of PLR. The structure of PLR was characterized by $^1$H NMR, XRD, and FTIR. The $^1$H NMR spectrum of PLR-30% is chosen as a typical example. As shown in **Figure 1b**, the signal at 3.50 ppm is assigned to the hydrogen atoms of the PEO axis and all the other peaks belong to the protons from *α*-CD rings. [27] The solubility of α-CD in PLRs in DMSO is restrained to a certain extent by PEO main chain, resulting in a relatively weak signal in $^1$H NMR spectrum. It can be seen from the $^1$H NMR curve that the ratio of $I_{PEO}$ (integrated area of the peak at 3.50 ppm) to $I_{Ha}$ (integrated area of the Ha signal in α-CD) is 1:2, which is significantly lower than the corresponding indicator (1:3, **Figure S1**) of the PEO/α-CD mixture. This significant difference shows that the PLRs is not a pure blend status; combined with many existed reports on synthesis of PLR based on PEG and α-CD, the successful formation of PLR can be proved. The number of binding sites on the PEO chain considering the fact that two ethylene glycol moieties are expected to be occupied by one α-CD ring (**Figure S2**). [28] Accordingly, the mass ratio between *α*-CD and PEO for saturated complexed PLR can be calculated as ~1100 wt.%. In this work, we choose 10 wt.%~80 wt.% of α-CD for the modification, aiming to select the best proportion (keeping the high latent heat and form stability) while keeping the essence of PEO main matrix. FTIR analysis was further employed to assess chemical structure among PLRs. **Figure 1c** shows the FTIR spectra of PEO/α-CD mixture and PLR-30 %. In curve PLR-30%, the peak at 961 cm$^{-1}$ is the stretching vibrations of C–H and the peak at 2883 cm$^{-1}$ belongs to –CH$_2$ of PEG. [29] The peak at 1020 cm$^{-1}$ and 3346 cm$^{-1}$ are the vibration of C–O–C and –OH in α-CD, respectively. It's further found that $I_{OH, PLR}/I_{CH_2, PLR}$ (It means the intensity ratio of -OH signal and -CH$_2$-) is much higher than $I_{OH, mixture}/I_{CH_2, mixture}$ (Figure 1c), which is because the signal of α-CD on the surface can be detected easily in PLR. While α-CD in the matrix may be restricted by the surface PEO, giving rise to weaker $I_{OH, mixture}/I_{CH_2, mixture}$. **Figure 1 d** showed that with the increase of α-CD content, the signal relative intensity of α - CD was enhanced (signal in green circle). Furthermore, based on the reports elsewhere [30] and the size analysis (**Figure S3**), we can conclude that only single PEO molecule chain can penetrate to the cavity of α – CD.

**Figure 1e** presents XRD patterns of the PLR films obtained for the film surfaces in reflection mode. The two strong peaks at 2θ angles of ~19° and ~23° for the PEO film are attributed to the (120) and (032) planes, respectively, of PEO crystals. As shown in **Figure 1e**, a significant α-CD diffraction peaks appear (the XRD curve and characteristic peaks of α-CD are given in **Figure S4**). There is no α-CD related diffraction peak in PLR-10% and PLR-30%, indicating that almost all of the α-CD forms the PLR structure in sample PLR-10% and PLR-30%. When the α-CD content is higher than 50%, the diffraction at 2θ=11.9° (110) and 7.6° (100) in sample PLR-80% appeared, which were most probably obtained from the PLR crystal consisting of α-CD and PEO. [31]

The DSC (**Figure S5**) results show that the melting point of PLR is lower than that of PEO. It is further found that a second stage of melting behavior appears in the high temperature region for sample PLR-50% and PLR-80%. Specifically, sample PLR-50% starts to melt at 98.32 °C, and sample PLR-80% starts at 106.18 °C. The melting behavior is consistent with the XRD results in **Figure 1h**. The melting point of the PEO crystals during the heating process shifted from 69.30 °C to 60.12 °C by compositing with $\alpha$-CD, and the corresponding crystallization point shifted from 41.44 °C to 46.15 and then further decrease to 39.31 °C. The melting point (Tm), latent heat of fusion (ΔHm), solidification temperature (Ts), and crystallinity ($\varphi_c$) of the pure PEO film and PLR films were determined by using DSC curves and listed in **Table S1**. With the introduce of α-CD, the crystallinity of the material just decreased slightly. This is mainly due to the relative freedom of PEO movement in the slide ring. According to the DSC curve of pure α-CD (**Figure S6**), there is no thermal change in the range of 50-150 °C. This shows that the thermal change around 100 °C is not derived from the pure crystals of α-CD. We speculate that this is the crystalline melting signal of α-CD in the PLR (PLR-α-CD). With the increase of a-CD, the crystallinity of α-CD on the PLR is relatively increased, so that $T_m$ of α-CD in the PLR increased.

We conducted solubility experiments on PLRs to conveniently compare the solvent resistance (**Figure 1f**). The solubility of PLR should be influenced significantly because lots of -OH is wrapped around the PEO chain. It is found that DCM can't completely dissolve all the PLRs. Furthermore, with the $\alpha$-CD content of PLR increasing, the degree of solidity increases. Based on the following two assumptions: (1) all CD participate in the formation of polyrotaxane structure (this assumption is based on the fact that no obvious α-CD diffraction signal of blending is detected in XRD), (2) and the formed polyrotaxane is insoluble in DCM, we can estimate the quantitative results of PLR formation by the insoluble content in Figure 1f. The specific results are shown in Figure S7. Specifically, the calculated PLR% indexes of PLR-10%, PLR-30%, PLR -50% and PLR -80% are 20.33%, 30.85%, 41.97% and 52.34%, respectively. From the perspective of practical application, we can directly use the feeding ratio to express the relative content of PEO. However, in practice, PLR with less cyclodextrin content on polyrotaxane may also be removed by DCM to a certain extent. Considering the possible existence of this part, the actual PLR content should be slightly higher than the above calculation index.

We can see that the material is PLR and some PEO raw materials which are not fully involved in the PLR. The schematic diagram of its micro molecular structure is shown in **Figure 1h**, **1i**, to **1j**. **Figure 1h** shows the PLR with low α-CD content such as PLR-10%. The vast majority of α-CD exists in the PLR form. When the content of α-CD is high, α-CD on the PEO chain will crystallize to a certain extent (Figure 1g). **Figure 1j** shows the microstructure illustration of the PEO/α-CD mixture. Since α-CD in PEO/α-CD mixture is weakly inhibited by PEO, α-CD in mixture is more prone to phase separation and crystallization, which gives rise to the remarkable signals in XRD curves.

*2.2 Optical properties and mechanical performance*

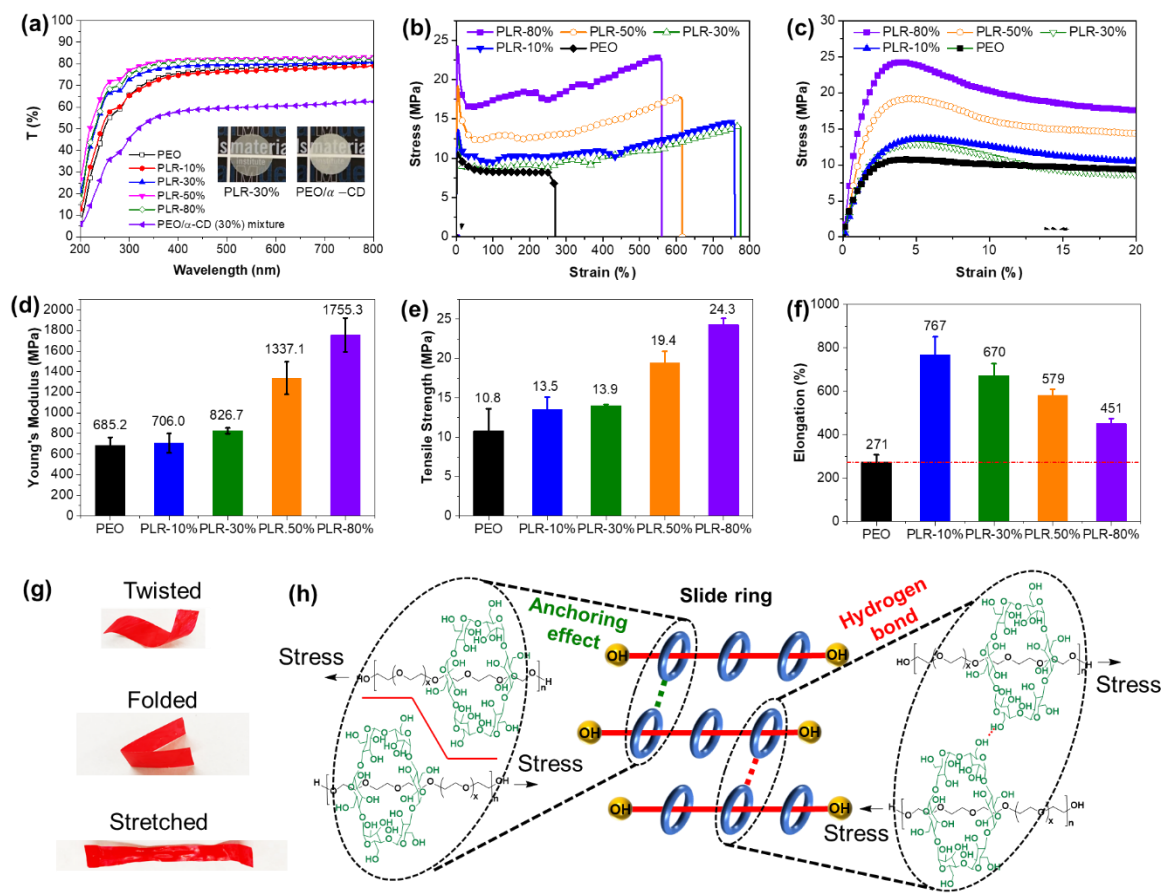

**Figure 2**. (a) Photographs of sample PLR-30% (Inset) and the PEO/$\alpha$-CD mixture and UV-Vis curves of all the samples (with 0.1 mm in thickness), (b) stress–strain curves, (c) stress–strain curves with local magnification, (d) Young's moduli (e) tensile strengths, and (f) elongations at break of the samples, (g) the images of twisted, folded and stretched sample (PLR-30%, please note the sample (PLR-30 %) was dyed with red pigment and the three images were obtained based one specimen), and (h) illustration of the mechanical enhancement mechanism.

UV-Vis spectra were used for the detecting transmittance performance of the PLRs. **Figure 2a** presents the UV–vis light transmission spectra of the PLR films and photographs of the pure PEO and PLR films. The PLR film had a high transmittance of ~80% at 400-800 nm wavelength, which is quite close to that of PEO. It is maybe because there is no significant phase separation in the PLR matrix due to the chemical structure of PLR. While the PEO/$\alpha$-CD mixture film was translucent and had a transmittance of only ~60 % at 400-800 nm wavelength.

Stress–strain curves of the PLR films are all presented in **Figure 2b** and **2c**. Detailed mechanical property data are summarized in **Table S2** and **Figure 2d-f**. The material has obvious yielding point during the tensile process, and the tensile strength of the material is the yield strength of the material. We further find that with the increase of α-CD content, the tensile strength and Young's modulus of the material increase gradually. Specifically, Young's modulus, tensile strength, and elongation at break of PEO are 685.2 MPa, 10.8 MP and 271 %, respectively; Young's modulus of PLR-80% is 1755.3 MPa (2.56 times of original PEO); tensile strength is 24.3 MPa, (2.25 times of original PEO), and the elongation at break increased to 451 %. After the modification, all the Young's modulus, tensile strength, and elongation at break of the PLR were significantly improved, indicating that the PLRs all have ultra-high flexibility (**Figure 2g**, it is clearly depicted that the samples can be twisted, folded, and stretched). **Figure 2h** is used to explain the specific reasons for the simultaneous realization of reinforcement and toughening of PLR. The interactions among α-CD in PLR include hydrogen bonding between α-CD and riveting among α-CD (specifically, steric hindrance mainly due to ). In the network, the relatively static α-CD units can act as the physical entanglement effect. Slide rings can ensure the free slip of PEO molecular chain, [23] which is equivalent to extending the length of molecular chain, and effectively avoiding stress concentration in the process of deformation, thus strengthening and toughening can be realized at the same time. Another possible reason for the mechanical enhancement is PLR considering PEO and α-CD may crystallize. The crystal can act as a physical crosslinking point in the material system.

The mechanical properties showed that the elongation at break of α-CD/PEO blends was reduced from 271% for pure PEO to 60% for PEO/α-CD (30 wt. %) blends (**Figure S8**). The significant difference in apparent properties between blends and PLR-30% is also an evidence to prove the formation of PLR structure. Considering the significant side effects of blending effect and the main topic in this work, the blending samples will be not further discussed and only the pure PEO samples are used as a reference for further characterization of the PLRs.

*2.3 Thermal performance*

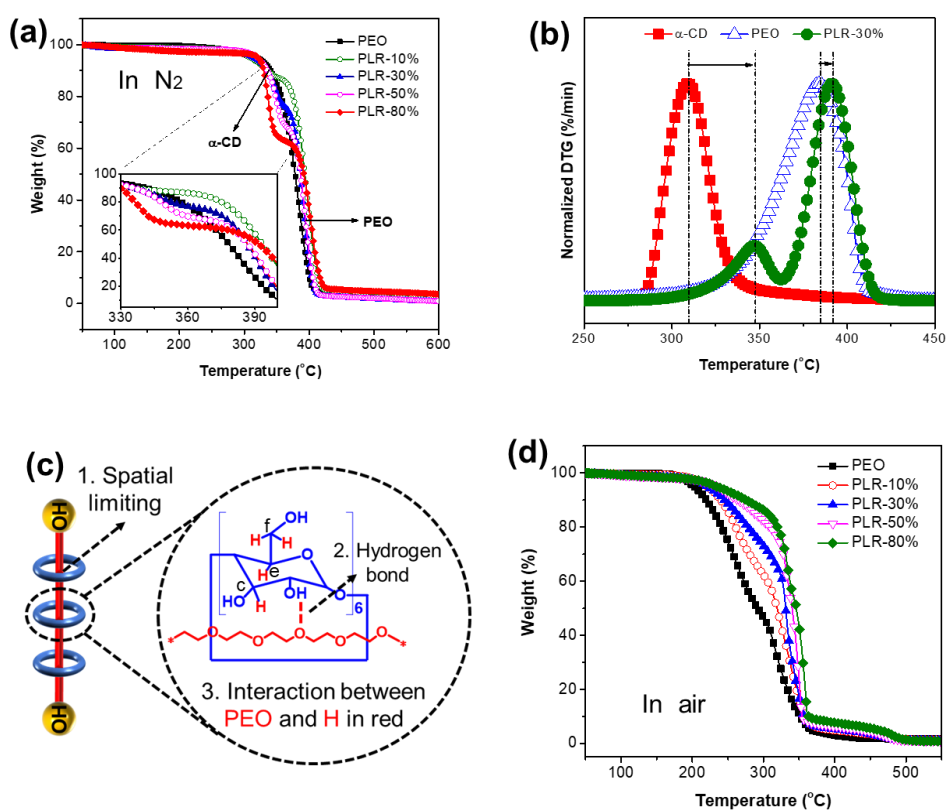

**Figure 3**. (a) TGA curves obtained in $N_2$ atmosphere, (b) typical DTG curves (in $N_2$) of samples α-CD, PEO and PLR-30%; (c) illustration of the interaction between the PEO thread and the $H_{c, e, f}$ (as marked in red) of α-CD [23] DTG of other samples are shown in Figure S9; and (d) TGA curves obtained under air atmosphere.

Thermo gravimetric analysis (TGA) and derivative TGA (DTG) analyses were carried out to determine the thermal stability of the PCMs. The corresponding results are presented in **Figure 3a** and **3b**, and **Table S3**. As it can be seen, all the PCMs underwent a two-step degradation process. The first step involved the degradation of the α-CD at ~350 °C. The second step occurred at approximately above 380 °C, which corresponds to the decomposition of the PEO chain.

According to Rajzer, [32, 33] an improved thermal stability could be achieved for the component with a lower $T_{max}$ if the measured $T_{max}$ of one component shifted to the higher $T_{max}$ of another component due to the various interactions between two components. In PLR materials, lower $T_{max}$ of $\alpha$-CD (309.1 °C) had shifted towards higher $T_{max}$ (**Figure 3b**, ≥332.3 °C) that indicated some interactions between PEO and $\alpha$-CD in the materials. In fact, the interaction can include the anchoring effect of PEO on α-CD, hydrogen bond between –OH in $\alpha$-CD and –O– in PEO, and the interaction between PEO and hydrogen elements, as shown in **Figure 3c** and well explained in the literature elsewhere. [23] This also leads to the decomposition temperature of PEO in PLR higher than that of pure PEO. We also analyzed the thermal decomposition behavior of the samples in air atmosphere. The results are shown in **Figure 3d**. It can be seen that the maximum thermal decomposition temperature increases with the increase of α-CD contents. Typically, when the content of cyclodextrin was 80 wt.%, the maximum thermal decomposition temperature increased by nearly 40 °C. The specific TGA in air of the materials are also listed in **Table S3**. These results show that the PCMs exhibit remarkably enhanced thermal stabilities in air than that of pure PEO.

*2.4 PCMs performance*

*2.4.1 Form stability*

**Table 1**. Some PCMs parameters

| Samples | Density (g cm$^{-3}$) | Melting Enthalpy (J g$^{-1}$) | Enthalpy efficiency (%) | Activation energy (kJ mol$^{-1}$) | Extent of supercooling (°C) | Heat lose (%) | Leakage | Form stability Size remaining at 80 °C (%) | Size remaining at 100 °C (%) |
|---|---|---|---|---|---|---|---|---|---|
| PEO | 1.33±0.05 | 115.30 | 100.00 | 228.48 | 8.95 | 5.46 | Yes | 76.17 | 63.25 |
| PLR-10% | 1.41±0.03 | 104.29 | 99.50 | 752.35 | 10.93 | 0.53 | No | 89.09 | 81.60 |
| PLR-30% | 1.42±0.07 | 88.57 | 99.87 | 571.07 | 11.44 | 2.07 | No | 98.08 | 97.77 |
| PLR-50% | 1.47±0.08 | 71.33 | 92.79 | 706.76 | 11.49 | 0.91 | No | 98.14 | 91.17 |
| PLR-80% | 1.56±0.04 | 57.11 | 94.11 | 533.07 | 9.51 | 0.38 | No | 98.22 | 98.00 |

Note: the extent of supercooling and heat lose were both calculated according to the first cycle in DSC results.

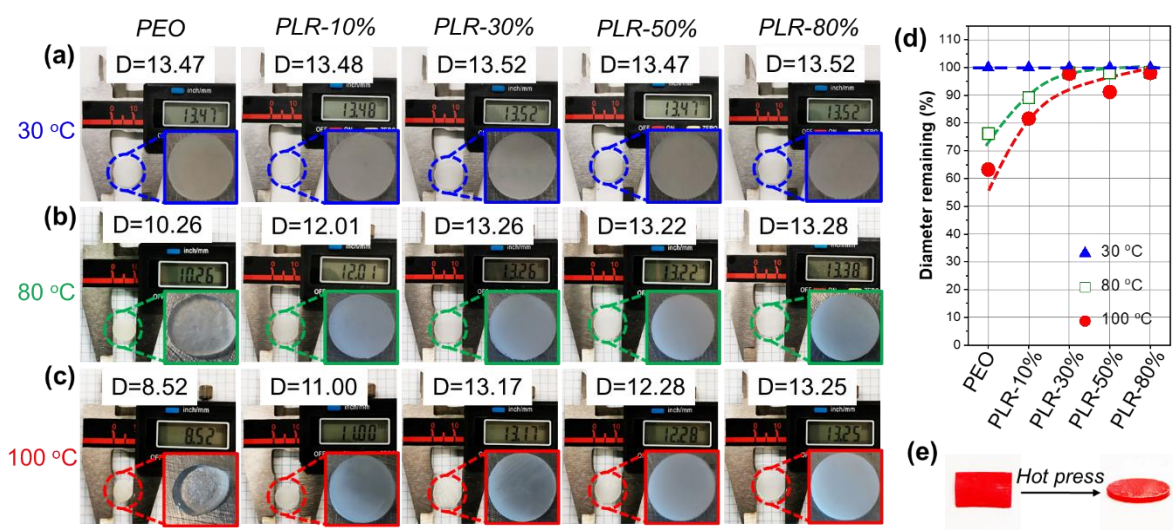

**Figure 4**. Form stability of the samples: diameter (D, mm) change of pure PEO and the four PLRs at different temperatures: (a), 30 ºC, (b) 80 ºC for 2 h, and (c) 100 ºC for 2 h, (d) change of sample diameter before and after heating treatment. (The diameter remaining can be calculated by the equation: $Diameter\ remaining\ \% = \frac{D_T}{D_i} \times 100\%$, where $D_T$ (mm) is diameter after heat treatment, and $D_i$ (mm) is the initial diameter), and (e) preparation of sample for the form stability test by hot press from the solvent casted sample (please note the sample was dyed with red pigment).

The photos of pure PEO and PLRs using hot plate treatment at different temperature are shown in **Figure 4a-c**. PEO gradually began to melt when the temperature reached the phase-transition temperature, and fully melted at 80 °C, but all the PLRs showed no significant changes in appearance, especially for the sample diameter. In addition, no leakage was observed during the entire heating process even when the temperature reached 100 °C, which was much higher than the phase-transition temperature of PEO ($T_{m,\ peak}$=63.9 ºC). These results indicate that the PLRs have excellent form stabilities. As it can be seen in **Figure 4d**, when the content of α-CD is higher than 30 wt.%, the diameter of the sample keeps higher than 91% (relative to the original diameter) even at 100 ºC (the specific data are listed in **Table 1**). This is mainly because when α-CD is introduced, the micro crystallization or micro aggregation of α-CD plays a role in the physical crosslinking point, so that the shape of the sample can be greatly maintained (as illustrated in **Figure 1g**). **Figure 4e** presents the sample preparation for the form stability by hot press. Due to the thermoplastic and flexible essence, the PCMs can be remolded conveniently.

*2.4.2 Thermal, storage and cycling performance*

The activation energy (ΔEa, kJ mol$^{-1}$) is related to the nucleation and mobility process of PCM molecules, and reflects the energy barrier for the phase transition of PCMs under the nonisothermal environment. It is reported that the ΔEa consists of nucleation activation energy that is the energy of forming the critical-size crystal nuclei, and transport activation energy that is the energy of transporting PCM molecular segments across the phase boundary [34].

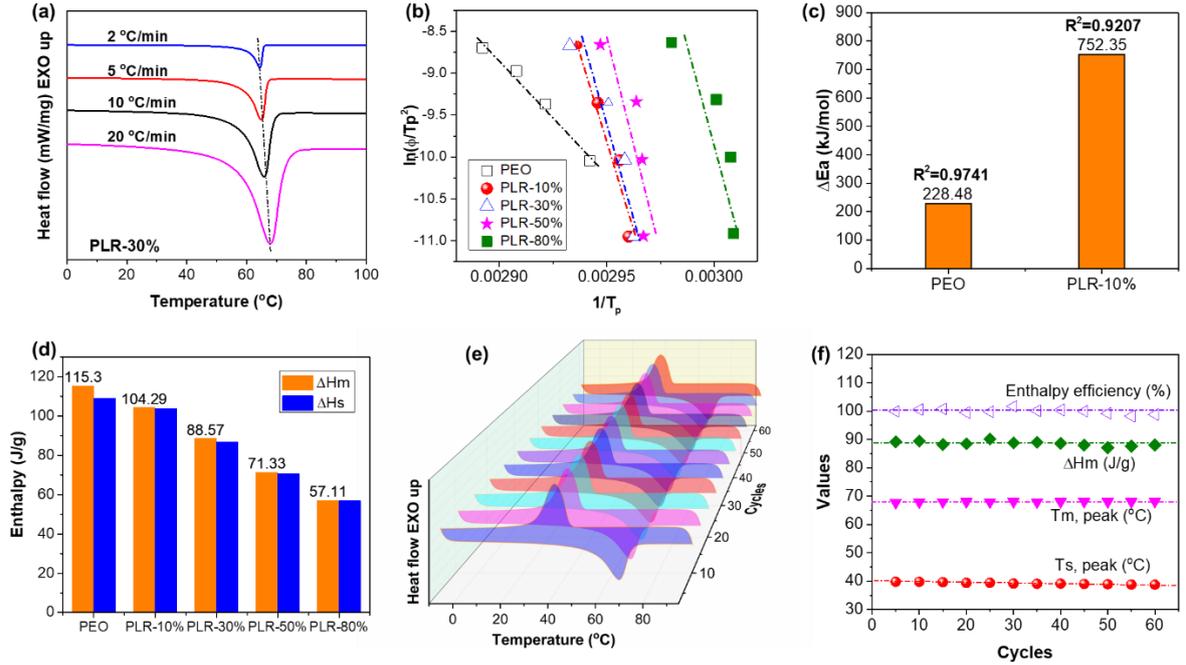

**Figure 5.** (a) Nonisothermal DSC curves of sample PLR-30% with different heating rates (2 °C min$^{-1}$, 5 °C min$^{-1}$, 10 °C min$^{-1}$ and 20 °C min$^{-1}$), (b) ln($\varphi$/Tp$^2$) versus 1/Tp plots; (c) $\Delta E_a$ of PEO and PLR-10%; (d) melting and solidification enthalpies of PCMs with different $\alpha$-CD contents at heating rate of 10 °C min$^{-1}$; (e) DSC cycle curves of PCMs (sample PLR-30%); and (f) the cycle performance (including the parameters of enthalpy efficiency, latent heat, melting temperature and solidification temperature) of PLR-30% as a typical example.

The $\Delta E_a$ of nonisothermal phase transition can be calculated by Kissinger's Equation:[34]

$$\frac{d[ln(\varphi/T_P^2)]}{d(\frac{1}{T_P})} = -\frac{\Delta E_a(T)}{R} \tag{1}$$

where $\varphi$ (K min$^{-1}$) is the heating rate, $T_P$ (K) is the phase change temperature and R (8.315 J mol$^{-1}$ K$^{-1}$) is the gas constant. $\Delta E_a$ can be calculated through the slope of the curve of $ln\left(\frac{\varphi}{T_P^2}\right)$ against $\frac{1}{T_P}$.

**Figure 5a** shows the nonisothermal melting peaks of PLR-30% determined at different heating rates. The DSC curves of other samples are shown in **Figure S10-13**. **Figure 5b** presents the Kissinger plots based on the data corresponding to **Figure 5a** and Figure S7-10. From the slope of the straight lines in the plots, the following values of the activation energy were obtained: E$_{a,PEO}$=228.48 kJ mol$^{-1}$ for neat PEO, E$_{a, PLR-10\%}$=752.35 kJ mol$^{-1}$ for PLR-10%. The activation energies for other samples are all listed in **Table 1**. We found that all the $\Delta E_a$ of the PLRs are much higher than that of pure PEO, which is mainly because the α- CD ring will restrain the mobility process of PEO chain in PCM to a certain extent.

The extent of supercooling (ΔT, °C) was evaluated by the following equation:

$$\Delta T = T_{m,onset} - T_{s,onset} \tag{2}$$

According to equation (2), the extent of supercooling of the material is calculated to be ~10 °C (Table 1). The undercooling degree did not change a lot, which indicated that the introduction of α- CD did not significantly hinder the nucleation process of PEO molecular chain. Moreover, the crystallinity degree of PEO is not significantly affected by the addition of $\alpha$-CD (**Table 1**), which is attributed to the relative sufficient movability of PEO due to the slide ring. PEO can slip freely in α-CD ring and α-CD did not significantly inhibit the chain folding of PEO. It is reported that the phase transition process of pure PEO is dominated by homogenous nucleation and growth mechanism. While, $\alpha$-CD can provide numerous heterogeneous nucleation sites for the phase transition of PEO. Normally, phase transition kinetics of PCMs are mainly determined by two factors when $\alpha$-CD are introduced [35]. On the one hand, $\alpha$-CD can serve as intramolecular heterogeneous nucleation agents during the phase transformation process of PCMs, thus yielding a positive effect. On the other hand, $\alpha$-CD can hinder the free mobility of PEO molecular chains and constrain the crystalline growth increase the transport activation energy (**Figure 5a-c**).

Furthermore, some parameters can be calculated for deeply understanding the PCMs, [36] the enthalpy efficiency of PCMs can be determined by equation 3, [37]

$$Enthalpy\ efficiency\ \% = \frac{\Delta H_m}{\omega \Delta H_{PCM}} \times 100\ \% \qquad (3)$$

where, $\Delta H_m$ (J g$^{-1}$) was the latent heat value of the PCMs. $\Delta H_{PCM}$ (J g$^{-1}$) represents the latent heat of PEO, and $\omega$ (%) represents the mass ratio of PEO in the PCMs.

The percentage of heat lose ($\eta$, %) was evaluated by equation 4:

$$\eta = \frac{\Delta H_m - \Delta H_s}{\Delta H_m} \times 100\ \% \qquad (4)$$

where, $\Delta H_m$ (J g$^{-1}$) was the latent heat value of the PCMs, and $\Delta H_s$ (J g$^{-1}$) is solidification enthalpy.

The enthalpy efficiency, heat lose and $\Delta H_m$ were also listed in **Table 1**. As shown in Table 1, the PCMs exhibited a relatively high-phase change enthalpy (57.11~104.29 J g$^{-1}$) (**Figure 5d**), high enthalpy efficiency, and almost unchanged extent of supercooling. The percentage of heat loss for pure PEO between endothermic and exothermic cycles was quite low (<2.07%), which indicated α-CD will not induce the heat loss of PEO in PCMs. The PCMs films had clear melting and crystallization temperatures in the heating and cooling processes, respectively, and had high cycle performance because the main parameters for PCM, such as $\Delta H_m$, $\Delta H_s$ and melting temperature, et al., remained virtually unchanged (**Figure 5e, f,** and **Table S4**) after 60 cycles. The PLR films are therefore regarded as high-performance form-stable phase change materials for thermal energy storage.

*2.4.3 Temperature response behavior*

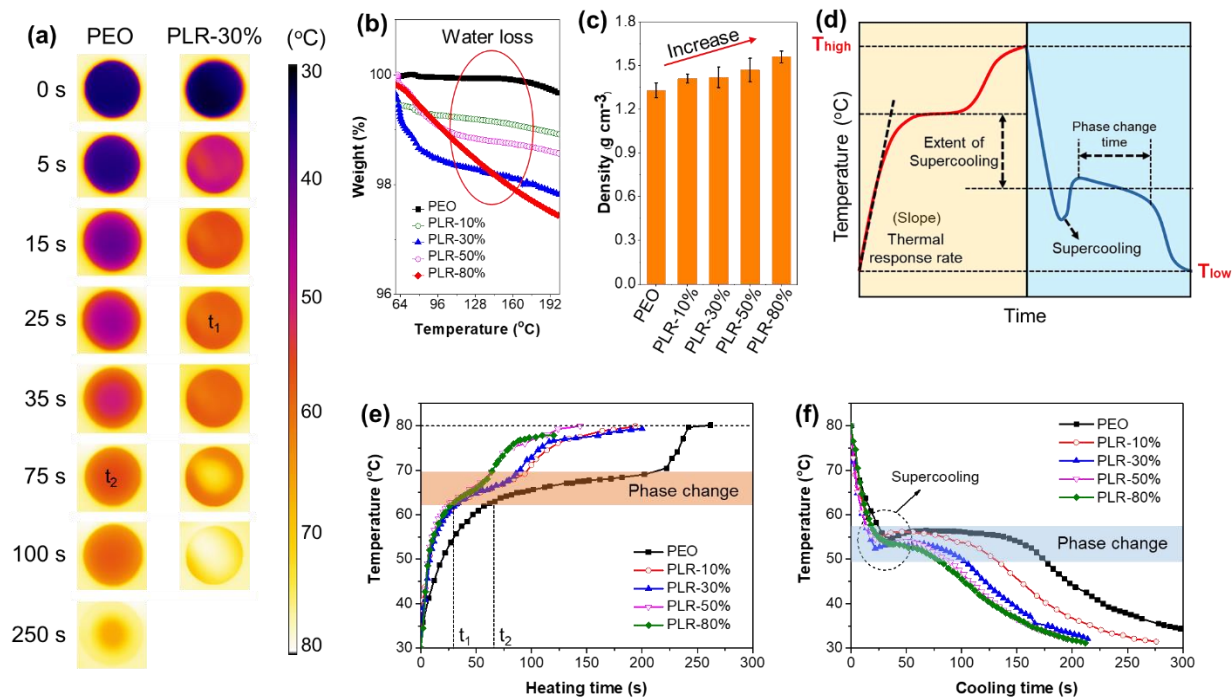

**Figure 6**. (a) Thermal images of pure PEO (left) and PLR-30% (right) during heating, (b) TGA curves with the range from 60 °C to 200 °C, (c) Density results of the PEO and PLR samples, (d) illustration of PCM curves during heating and cooling, (e) the heating curves in which $t_1$ and $t_2$ mean the time from the beginning of the heating of the PCMs and PEO, respectively, to the beginning of the phase change, and (f) cooling curves of the PEO and PLR samples.

The temperature response behaviors of the pure PEO and PCMs were further examined using an infrared camera to evaluate their thermal response rates and thermal transfer capacities, as shown in **Figure 6**. During heating process, it can be clearly seen that the color of PCM (PLR-30% was chosen as a typical example and the images of rest of the samples are provided in **Figure S14**) changed much faster than that of pure PEO from the temperature distribution images at 0, 5, 15, 25, 35, 75, and 100 s, indicating that the thermal response rate of the PCMs is much faster than pure PEO. As shown in Figure 6b, the TGA results showed that there could be 1wt.%~2wt.% of absorbed water in the PCMs. Meantime, Figure 6c depicted that the density of the PCMs was increased with the increase of $\alpha$-CD contents (see Table 1 for the exact density values). It is well reported that the increase of water content will cause the increase the thermal conductivity. [38-40] Moreover, increasing bulk density decreases the average distance among molecules, transfers its energy to a different molecule conveniently, and increases thermal conductivity accordingly. [41] Therefore, we here assigned that the increasing heating rates of the PCMs was most likely due to both the increased water contents and increased density. Accordingly, the PCMs can respond quickly to changes in the temperature of the external environment and exhibit a good, intelligent temperature-control effect. These results illustrate that the novel PCMs have a great potential application for thermal energy storage systems with fast thermal response rate.

To clearly evaluate the improvement in the thermal-regulating behavior of the PCMs, temperature-time curves of pure PEO and PCM were investigated in a temperature range from 30 °C to 80 °C as seen in **Figure 6e** and **6f**. Notably, **Figure 6d** illustrates the regular curves with clear explaining of each part. It is evident that the temperature of both pure PEO and PCM takes as long as 100 s~250 s reaches a thermal balance of 80 °C during the heating time. This indicates that the interior PEO of the PCMs can be fully molten and absorb the exterior heat through phase change procedures to maintain their temperature. The thermal response rates (the slope of the curves based on the most beginning) of the PCMs are significantly higher than that of pure PEO. Therefore, it's obviously found that $t_1$ was lower than $t_2$ (Figure 6a and 6e). It can be distinguished that the temperature of PEO and PCM tend to be constant in the temperature range of 60 °C through 70 °C, which is near their phase change temperature range. The total heat procedure lasts about 250 s for PEO, while it lasts about 150 s for all the modified PCMs. The appearance of this shorter temperature plateau of PLRs may be due to both low latent heat and high thermal response rate (as mentioned above). The same results are also present in the cooling temperature-time curves as illustrated in **Figure 6d**. The temperature of the pure PEG and PCM can be maintained in the range between 55 °C and 58 °C for approximately 150 s for the PEG and 70 s (phase change time) for PLR-30%. These results demonstrate that the PCMs have a potential application in fast thermal management for both electronic devices and thermal energy storage systems. Furthermore, the extent of supercooling is ~10 °C, which is in good agreement with that obtained based on DSC results.

2.4.4 Shape memory properties

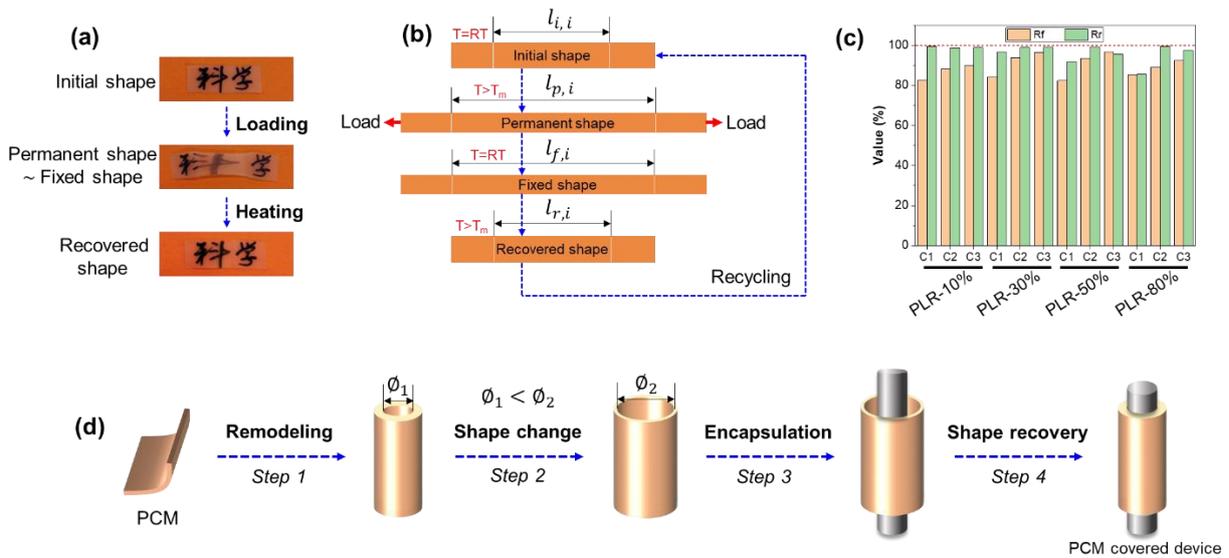

**Figure 7**. (a) Images of material (with the words of "Science" in Chinese) deformation and recovery, (b) schematic diagram of shape memory test procedure, (c) the $R_f$ and $R_r$ testing results (3 cycles (C1, C2 and C3) for each sample. Please note that the samples used for shape memory test were produced by hot press and cut into rectangle shape), and (d) illustration of devices packaging with the novel PCMs by 4 steps.

The PCMs show good shape memory properties. **Figure 7a** shows the images of material (with the words of "Science" in Chinese) deformation and recovery. The examination of heat-induced shape memory properties focused on a uniaxial force of stretching to a specified percent elongation using the procedure described as follows (**Figure 7b**): (1) the sample (with length of $l_i$) was deformed at 80 °C with strain 250 % ($l_p$), (2) the sample was cooled to room temperature under constant tensile stress to freeze the extended strain ($l_p$). Subsequently, the applied stress was released, and the strain remains constant due to the restrain of crystalline domains ($l_f$); (3) the sample was heated up to 80 °C again, and the extended strain started to recover when the temperature exceeded $T_m$ of the sample ($l_r$). **Figure 7c** and **Table S5** showed the values for strain fixity ($R_f$) and strain recovery ($R_r$) for each sample within 3 cycles, namely, C1, C2 and C3. As it can be seen, the $R_f$ is with ~90 % of the initial strain after unloading of the stress, and the $R_r$ is with ~99 % of the strain recovering instantaneously. It should be mentioned that the PEO will completely molten at the test temperature. Therefore, the $R_f$ and $R_r$ cannot be provided. Notably, its ultra-high flexibility, remolding ability and excellent shape memory properties provide a convenient way for the intelligent heat treatment packaging of electronic devices or some other items (**Figure 7d** and **Figure S15**). Typically, the synthesized raw materials (sheet-like) can be remolded in to any other expected shape (Step 1). The initial shape (size) will be enlarged for the convenience of encapsulation or packaging (Step 2 and Step 3). The PCMs can be finally recovered by heating under proper temperature and coated on the surface of target devices (Step 4). It is finally worth pointing out that the PCMs can maintain good flexibility and mechanical strength above $T_m$, which is why it can exhibit excellent shape memory characteristics. This may expand the application of this type of PCMs to a certain extent. As far as we know, this has not been reported in other traditional PCM researches.

## 3. Methodology discussion

*3.1 Performance and preparation of PCMs*

**Table 2.** Information about PCMs based on PEG (or PEO) in the literature elsewhere.

| Composition | $M_n$ (kg mol$^{-1}$) | Methods | Solvent | $\Delta H_s$ (J g$^{-1}$) | $\Delta H_m$ (J g$^{-1}$) | Heat lose (%) | Enthalpy efficiency (%) | Shape memory | Ref. |
|---|---|---|---|---|---|---|---|---|---|
| PEG/Halloysite nanotube | 35 | Melt-extrusion | - | 97.0-115.8 | 96.8-115.9 | - | >95.16 | N.D. | [7] |
| PEO/CNC | 1000 | Solution mixing | $H_2O$ | - | 66–93 | - | - | N.D. | [19] |
| PEG/Si$_3$N$_4$ nanowires | 4 | Solution mixing | Ethanol | 123.6-131.8 | 139.5-152.3 | 11.4-13.4 | 92.27-96.93 | N.D. | [36] |
| PEG/cellulose acetate | 10 | Electrospinning | DMAc and acetone | 65.15 | 86.03 | 24.27 | - | N.D. | [42] |
| PEG/cellulose acetate/TDI | 10 | Electrospinning +crosslinking | Toluene, Acetone and DMAc | 25.85-65.15 | 36.72-86.03 | 24.27-29.60 | - | N.D. | [43] |
| PEG/TDI/Tetrabromobisphenol-A | 10 | Copolymerization | Toluene | 71.28-99.26 | 71.14-98.68 | - | - | N.D. | [44] |
| PVA/GA/PEG | 6 | Sol-gel method/hydrothermal reaction (95 °C) | $H_2O$ | 118.2-144.1 | 119.6-145.8 | <1.7 | - | N.D. | [45] |
| β-CD/MDI/PEG | 8 | Crosslinking copolymers | DMF | 95.27- 115.20 | 90.34- 111.60 | 3.13-5.17 | 69.02-88.01 | N.D. | [46] |
| Xylitol/MDI/PEG | 4 and 6 | Crosslinking copolymers | - | 64.25- 76.37 | 68.4- 80.46 | - | - | N.D. | [47] |
| Halloysite nanotubes/PEG/HDIB | 4 | Crosslinking copolymers | DMF | 83.5-120.6 | 83.8-123.7 | - | 76.6-82.9 | N.D. | [48] |
| Graphene oxide/PEG/ HDIB | 4 | Crosslinking copolymers | DMF | 71.6-76.3 | 71.7-78.0 | - | 52.47-57.32 | N.D. | [49] |
| Graphene oxide/PEG/MDI | 8 | Copolymerization | DMF | 134.3-146.1 | 136.0-146.1 | - | - | N.D. | [50] |
| PEG/IPDI/phloroglucinol | 8 | Crosslinking (PU) | Methyl ethyl ketone | 88.4-125.6 | 117.1-146.6 | 9.88-24.55 | - | N.D. | [51] |
| PEG/glucose/MDI | 8 | Crosslinking (PU) | DMF | 106.7-121.4 | 108.7-131.9 | - | - | N.D. | [52] |
| PEG/MDI/castor oil | 4 and 6 | Crosslinking (PU) | - | 72.39-109.00 | 76.63-117.70 | 5.53-7.39 | 57.05-71.67 | N.D. | [53] |
| PEO/α-CD | 900 | Host-guest recognition | $H_2O$ | 56.89-103.74 | 57.11-104.29 | 2.07-0.38 | 92.79-99.87 | $R_f$~90 %, $R_r$~99 % | This work |

**Note:** N, N-dimethylacetamide=DMAc; Hexamethylene Diisocyanate Biuret= HDIB; 4, 4'-diphenylmethane diisocyanate=MDI; N, N-dimethylformamide=DMF; Phosphorylated polyvinyl alcohol=PPVA; graphene aerogel=GA

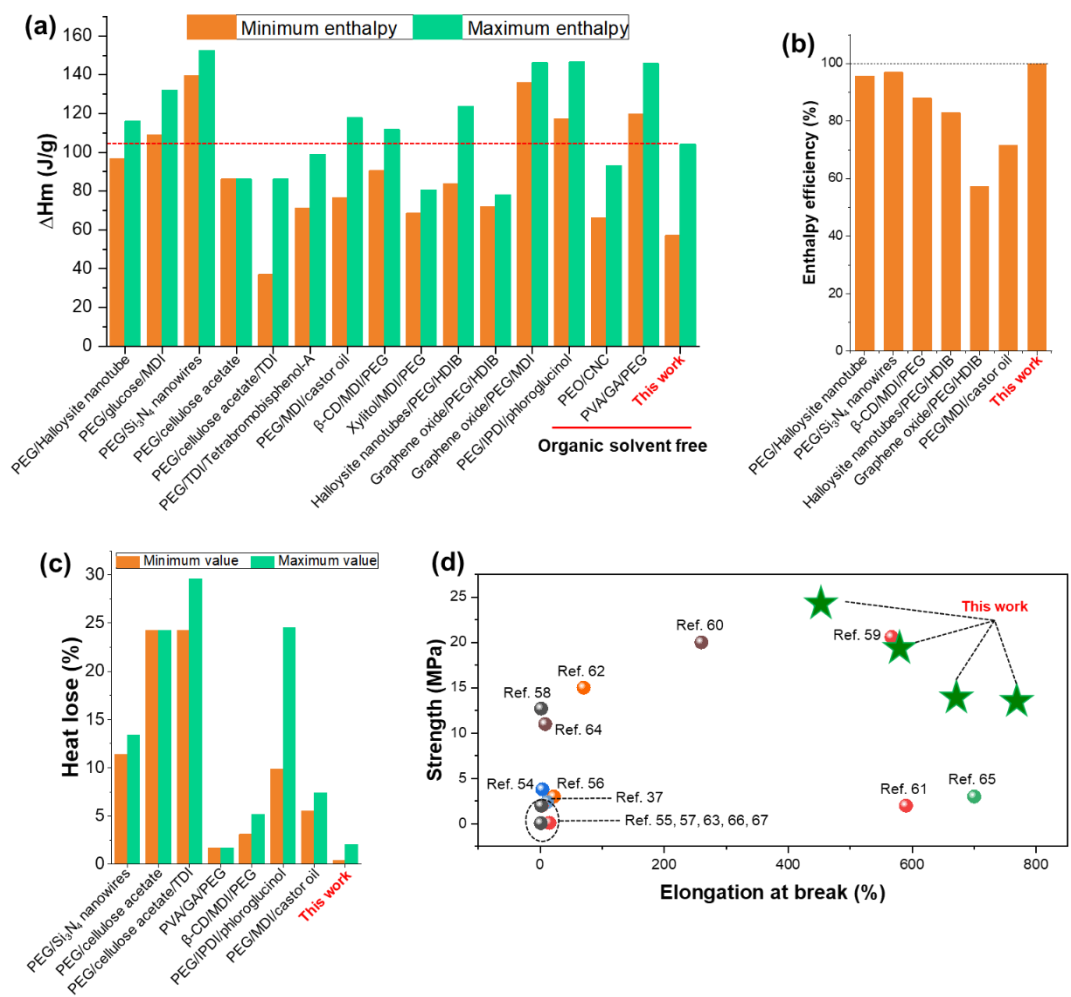

**Figure 8**. Compare for (a) latent heat, (b) enthalpy efficiency, (c) heat lose and (d) flexibility according to the reports elsewhere.

A comparison was made for the latent heat preparation method, heat lose, melting and solidification enthalpy, and enthalpy efficiency of pure PEG, the present PCMs and other PCMs reported in the literature to highlight the enhancement of the thermal performance, the PCM preparation method, and flexibility of the novel PCM (**Table 2**, and **Figure 8**). It is worth noting that the conventional PEG-based PCM are prepared by melt-extrusion, solution mixing, and copolymerization or crosslinking, and encapsulation by other 3D-frameworks. Organic solvents and petrol-based chemicals were usually used in most of the works. In contrast, the advantages of this study are also reflected in the green and convenient preparation process. This work reports the convenient and efficient preparation of ultra-flexible and shape stable PCMs. Therefore, the materials reported in this study have significant advantages in preparation methods and some properties.

As shown in **Figure 8a**, the latent heat is with the same high level as the reported in some other works. While the enthalpy efficiency is significantly higher than the reported ones (**Figure 8b**). The materials have extremely low heat lose (**Figure 8c**), which indicating the latent heat can be released during the cooling step and increase the energy efficiency to some extent. We further compare the mechanical properties with some reported flexible PCM (not only PEG) composites. [37, 54-67] As shown in **Figure 8d**, the present samples showed ultra-high strength and elongation at break. Notably, the ultra-high flexibility, mechanical strength and toughness of the material give us a great degree of freedom for further modification. In other words, we can achieve the improvement of some other core indicators, e. g. latent heat, by filling other components into the PLR matrix under the premise of ensuring the flexibility of materials.

*3.2 Method analysis*

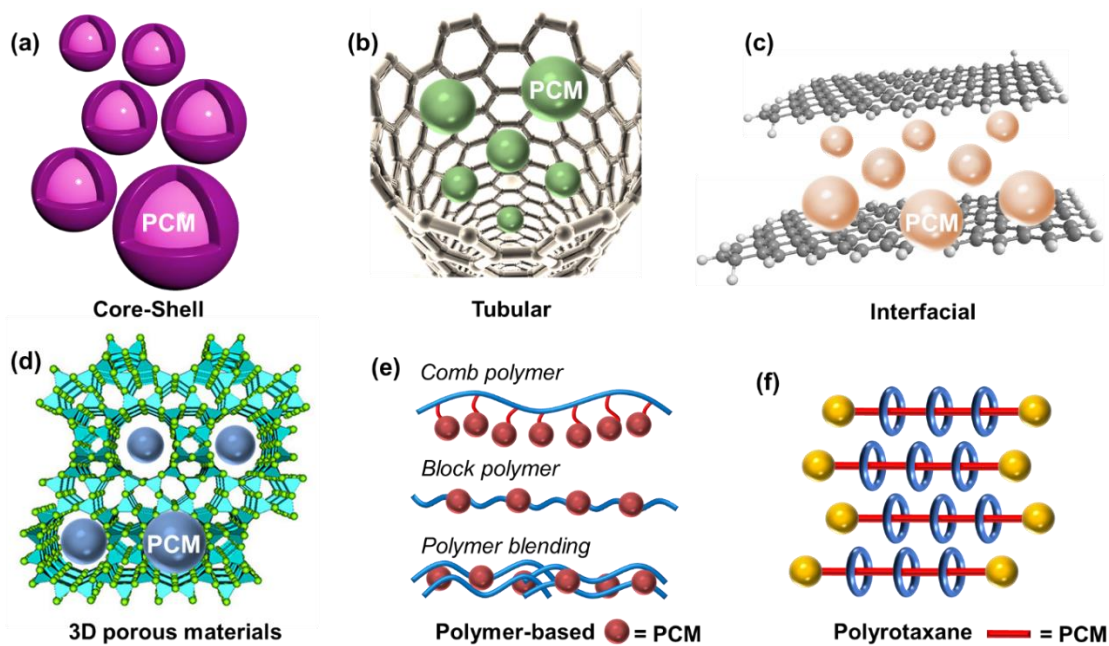

**Figure 9.** Methods for form-stable phase change materials.

At present, investigation of PCMs has focused on shape and strengthening heat transfer in order to solve application problems. Form-stable PCMs are composite materials comprising solid–liquid PCMs and support materials. The conventional fabrication methods include the encapsulation of PCMs in polymer matrices by blending or in-situ polymerization, (**Figure 9**a) PCM encapsulation in tubes (Figure 9b) or sheet-form nanomaterials, (Figure 9c) PCM injecting into 3D porous materials, (Figure 9d) and PCM grafting onto the skeletons or blending of high-melting-temperature polymers, generating the comb polymers and block copolymers. (Figure 9e) [68-71] Notably, as a new method (the PCM structure is illustrated in Figure 9f), it has the advantages as following: a) simple and eco-friendly method; b) outstanding performance: ultra-flexible, high strength, high thermal stability; c) sustainable: application of biomass $\alpha$-CD and biodegradable PEO; d) convenient performance control; e) convenient for re-processing and re-molding and f) excellent shape memory performance. It also suffers from two weakness: a) PLR varieties limitation, and b) limited phase change enthalpy. Fortunately, we can make improvement for the weakness mentioned in the work one by one: 1) expand other PCMs, e.g. PCL based PLR, 2) load additional PEG to improve the phase change enthalpy.

## 4. Conclusions

PEO based PLR with different $\alpha$-CD contents were successfully synthesized by using water as solvent via facile one-step method. Compared with the PEO, the thermal stability in the air and tensile properties of PLR were significantly improved. Typically, the young's modulus, tensile strength, and elongation at break of the PEO film were 685.2 MPa, 10.8 MPa and 271 %, respectively, and remarkably increased to 1755.3 MPa, 24.3 MPa, and 451 % respectively, with the addition of 80 wt.% of α-CD. The maximum thermal decomposition temperature in the air increased by nearly 40 ºC. Nevertheless, the form stability at temperatures above the melting point of PEO significantly increased with the α-CD addition. When the content of α-CD is higher than 30 %, the diameter retention rate is higher than 91 %. The PCMs possess outstanding shape-fixing and recovery properties (shape-fixing and recovery ratios are about 90 % and 99 %, respectively). Further with the relative high phase transition enthalpy (57.11-104.29 J g$^{-1}$), excellent cycle performance and fast heat response rate, the PLR films are therefore promising sustainable and advanced form stable phase change materials for energy storage. Notably, the typical advantages of this method are that PCM materials with high strength, high toughness and flexibility, high form stability and excellent shape memory properties can be prepared eco-friendly and efficiently. Due to the wide modification possibilities, PLR is expected to open a window to the research of high performance PCMs. In addition, the repeatable processability broadens the PCM applications in the field of smart heat treatment for chip or battery packaging.

**Declaration of Competing Interest**

The authors declare that they have no known competing financial interests or personal relationships that could have appeared to influence the work reported in this paper.